\documentclass[11pt]{article}
\usepackage{amsmath}
\usepackage{mathptmx}
\usepackage{amsfonts,longtable}
\usepackage{epsfig}
\usepackage{amsfonts}
\usepackage{color}
\def\R{\hbox{{\rm I}\kern-0.2em{\rm R}\kern0.2em}}

\def\be{\begin{equation}} \def\ee{\end{equation}} 
  \def\({\left(} \def\){\right)} 
\def\[{\left[}
\def\]{\right]}
\def\bc{\begin{center}}
\def\ec{\end{center}}

\begin{document}
{\bc {Noether Symmetries of Bianchi Type II Spacetime Metrics}\ec}
{\bc {Mahmood Rajih Tarayrah, Hassan Azad, Ashfaque H.
Bokhari,and~ F. D. Zaman }\ec} {\bc Department of Mathematics and
Statistics, King Fahd University of Petroleum \& Minerals, Dhahran
31261, Saudi Arabia\ec} {\bc ABSTRACT \ec} We classify a class of
Bianchi type II spacetimes according to their Noether symmetries.
We briefly discuss the conservation laws admitted by these Noether
symmetries and classify their possible algebras and determine
their structures also.\\
{\bf Introduction}\\ The Noether symmetry, also known as
variational symmetry, is associated with mechanical systems
possessing a Lagrangian $L$. In general Relativity, this
Lagrangian  is given by $L=g_{ab}{\dot x^{a}}{\dot x^{b}}$ and is
obtained from the spacetime metric $g_{ab}$ given by
$ds^2=g_{ab}dx^a dx^b.$ The Euler-Lagrange equations corresponding
to the above Lagrangian determine the trajectories of the
particles and are solutions of ${\ddot {x^a}}+\Gamma
^{a}_{bc}{\dot x^b}{\dot x^{c}}=0,$ also known as geodesic
equations. In Relativity, the spacetime metric, $g_{ab}$ is of
particular significance as it determines exact solutions of the
Einstein field Equations, $ R_{ab}-(1/2)g_{ab}R=\kappa T_{ab},$
where $R_{ab}, T_{ab}, R$ and $\kappa$ respectively represent
Ricci \&
Stress tensors, Ricci scalar and the gravitational constant.\\
As for Einstein, they are highly non-linear coupled system of
partial differential equations. This non-linearity makes Einstein
field equations very hard to solve for their exact solutions. On
the other hand if one chooses energy momentum tensor to define
Einstein field equation, then any arbitrarily chosen metric will
form a solution of the these equations, which may be totally
un-physical. To date reasonably small number of solutions of the
Einstein field equations are known and their full account can be
found in \cite {mtw,petrov}. Generally, exact solutions are
categorized as vacuum and non-vacuum solutions. One of the most
interesting solutions of the vacuum Einstein field equations is
given by the famous Schwarzschild solution, whose predictions
(planetary orbits and black holes etc) are responsible for the
fundamental importance attached to general theory of Relativity.
In the non-vacuum case, there are several classes of solutions
that have been widely discussed. Brans and Dicke \cite{brans}
theory of gravitation is a well known modified version of
Einstein's theory of Relativity. It is a scalar tensor theory in
which the gravitational interaction is mediated by a scalar field
$\phi$ as well as the tensor field $g_{ab}$\cite{uma}. In a recent
work Singh and Rai \cite{singh} list a detailed  discussion of
Brans- Dicke cosmological models. In particular, spatially
homogeneous Bianchi models in Brans-Dicke theory in the presence
of perfect fluid are quite important to discuss the early stages
of evolution of the universe. Other important aspects of the Brans
and Dicke theory are discussed in Belinskii and Khalatnikov \cite
{belinski}, and Reddy et al. \cite{reddy}. Chakraborty and Bali
et.al  discussed Bianchi type IV strings models in general
Relativity \cite{ chak, bali1,bali2}. A detailed discussion of
Bianchi type II, VIII and IX models  in scale covariant theory of
gravity is given in the work published by Reddy et.al \cite
{reddy2}. Shanthi et.al have studied Bianchi type-VIII \& IX
models in Lyttleton-Bondi Universe \cite {shanthi}. Also, Rao
et.al have studied Bianchi type-VIII \& IX models in Zero mass
scalar fields and self creation cosmology \cite {Rao}. More
recently Valagapudi studied Bianchi type-II, VIII and IX perfect
fluid cosmological models in a scalar tensor theory \cite
{vala}.\\ Bianchi spacetimes have also been discussed in
literature from the point of view of symmetry approach. Following
this course, Capozziello et. al have investigated homogeneous
theories of gravity, in which a scalar field is minimally coupled
to gravity, searching for point symmetries in the cosmological
Lagrangian density which allow to solve exactly the dynamical
problem used. In particular, they have used Noether symmetry
approach to study the Einstein equations minimally coupled with a
scalar field, in the case of Bianchi universes of certain classes
\cite{cap}. In an interesting recent work Camci et. al have
investigated matter collineations (MC) of Bianchi type II
spacetime according to the degenerate and non-degenerate energy
momentum tensor \cite{camci}. It is shown that when the
energy-momentum tensor is degenerate, most of the cases yield
infinite dimensional MCs whereas some cases give finite
dimensional Lie algebras in which there are three, four or five
MCs. For the non-degenerate matter tensor cases they found that
the Lie algebra of MCs is finite dimensional, in which the number
of MCs is three, four or five. Furthermore, they discussed the
physical implications of the obtained MCs in the case of perfect
fluid as source.\\Our main focus in this paper is to classify
Bianchi type II spacetimes according to their metrics and Noether
symmetries. These models are worth studying because they present a
"middle way" between Friedmann Robertson Walker models and
completely inhomogeneous and anisotropic universes and thus play
an important role in modern cosmological studies. Further, the
Bianchi type-II models have been widely studied for the
simplification and description of the large scale behavior of the
actual universe. With this in mind, it is argued that it might be
worth investigating the variational conservation laws admitted by
the Lagrangian of the Bianchi spacetimes.\\ The plan of this
investigations is as follows: In the next section we give a brief
description of the equations giving Noether symmetries. In section
3, we give a complete classification of the Noether symmetries of
Bianchi type II spacetime metric and then classify Lie algebras of
Noether symmetries. A brief summary and discussion
of the work is given in the last section.\\
{\bf Noether Symmetries in Bianchi Type II Spacetime}\\ The
Bianchi type models represent a "middle way" between Friedmann
Robertson Walker models and the completely inhomogeneous and
anisotropic universes and thus play an important role in modern
cosmological studies. In particular, the Bianchi type-II models
have been widely studied for the simplification and description of
the large scale behavior of the universe. Mathematically, these
models are represented by a metric of the form~\cite{camci},
\begin{equation}\label{metric}
ds^2 \,=  - dt^2  + A(t)^2 dx^2  + B(t)^2 (dy^2  + x^2~dz^2  - 2x
~ dy~ dz) + C(t)^2~dz^2.
\end{equation}
The Lagrangian associated with the above metric is,
\begin{equation}\label{e204}
L =  - \dot t^2  + A(t)^2~\dot x^2  + B(t)^2 (\dot y^2  + x^2~\dot
z^2  - 2x\dot~y~\dot z) + C(t)^2~\dot z^2,
\end{equation}
where the dot (.) represents derivative with respect to the affine
parameter\, $s$. According to Noether's theorem, to each
differentiable symmetry of the action of a physical system there
corresponds a conservation law. For the time translation symmetry,
Noether's theorem states that the energy of the system is
conserved. Similarly, a translation in space gives rise to
conservation of linear momentum, while existence of a rotational
symmetry provides conservation of angular momentum.
Mathematically, the Noether Symmetry arises from the invariance
properties admitted by the lagrangian $ \mathcal{L}=\int_\Omega
L(t(s),x(s),y(s),z(s);s)~ds$ defined over $\Omega$. Including
point dependent gauge functions $f$, the Noether symmetries
associated with the given lagrangian are given by
\cite{karabokhari}
\begin{equation}\label{e44} X^{1}L + LD_s \xi  = D_s f,
\end{equation}where \begin{equation}\label{pro}
X^{1} = X + \tau ^1 \frac{\partial }{{\partial \dot t}} + \xi ^1
\frac{\partial }{{\partial \dot x}} + \eta ^1 \frac{\partial
}{{\partial \dot y}} + \varphi ^1 \frac{\partial }{{\partial \dot
z}},\end{equation} is the $1st$ prolongation of the symmetry
generator $X=\mu \frac{\partial }{{\partial s}} + \tau
\frac{\partial }{{\partial t}} + \xi \frac{\partial }{{\partial
x}} + \eta \frac{\partial }{{\partial y}} + \varphi \frac{\partial
}{{\partial z}} $~associated with 4-spacetime,~ in which $ \mu
,\,\,\tau ,\,\,\xi ,\,\,\eta ,\,\,\varphi$\, are functions of
$s,\,t,\,x,\,y,$ and $z$. Also, in (\ref{pro}) the components
$\tau ^1,\,\xi ^1,\,\eta ^1,\,\varphi ^1$ of the $1st$
prolongation are defined by \begin{equation}
\begin{array}{l}
 \tau ^1  = D_s \tau  - \dot tD_s \mu , \\
 \xi ^1  = D_s \xi  - \dot xD_s \mu , \\
 \eta ^1  = D_s \eta  - \dot yD_s \mu , \\
 \varphi ^1  = D_s \varphi  - \dot zD_s \mu, \\
 \end{array}
\end{equation}
with
\begin{equation}
D_s  = \frac{\partial }{{\partial s}} + \dot t\frac{\partial
}{{\partial t}} + \dot x\frac{\partial }{{\partial x}} + \dot
y\frac{\partial }{{\partial y}} + \dot z\frac{\partial }{{\partial
z}},
\end{equation}
defining  total derivative operator. Using equation (\ref {e204}),
gives rise to a polynomial equation in derivatives of $(t, x, y,
z)$:
\begin{equation}
\begin{array}{l}
 [\mu \frac{\partial }{{\partial s}} + \tau \frac{\partial }{{\partial t}} + \xi \frac{\partial }{{\partial x}} + \eta \frac{\partial }{{\partial y}} + \varphi \frac{\partial }{{\partial z}} + \tau ^1 \frac{\partial }{{\partial \dot t}} + \xi ^1 \frac{\partial }{{\partial \dot x}} + \eta ^1 \frac{\partial }{{\partial \dot y}} + \varphi ^1 \frac{\partial }{{\partial \dot z}}][ - \dot t^2  + A(t)^2 \dot x^2  +  \\
 B(t)^2 (\dot y^2  + x^2 \dot z^2  - 2x\dot y\dot z) + C(t)^2 \dot z^2 ] + [ - \dot t^2  + A(t)^2 \dot x^2  + B(t)^2 (\dot y^2  + x^2 \dot z^2  - 2x\dot y\dot z) +  \\
 C(t)^2 \dot z^2 ][\frac{\partial }{{\partial s}} + \dot t\frac{\partial }{{\partial t}} + \dot x\frac{\partial }{{\partial x}} + \dot y\frac{\partial }{{\partial y}} + \dot z\frac{\partial }{{\partial z}}]\mu  -[\frac{\partial }{{\partial s}} + \dot t\frac{\partial }{{\partial t}} + \dot x\frac{\partial }{{\partial x}} + \dot y\frac{\partial }{{\partial y}} + \dot z\frac{\partial }{{\partial z}}]f
=0
 \end{array}
\end{equation}
Setting the coefficients of $\dot t^3$,~$\dot x^3$, ~$\dot y^3$,
~$\dot z^3$, ~$\dot t^2$, ~$\dot x^2$, ~$\dot x^2$, ~$\dot z^2$,
~$\dot t\dot x$, ~$\dot t\dot y$, ~$\dot t\dot z$, ~$\dot x\dot
y$, ~$\dot y\dot z$, ~$\dot t$, ~$\dot x$, ~$\dot y$, ~$\dot z$
and the constant term, respectively, to zero, gives rise to an
over determined system of differential equations given by:
\begin{flalign}
&\mu~_t= 0,&\label{e56}\\
 &A^2\mu_x=0,&\label{e57}\\
 &B^2\mu_y =0,&\label{e58}\\
&(C^2+ B^2 x^2 )\mu _z = 0,&\label{e59}\\
&\mu~_s  - 2\tau~ _t= 0,&\label{e60}\\
&2AA'\tau  + 2A^2 \xi~ _x  - A^2 \mu~ _s = 0,&\label{e61}
\end{flalign}
\begin{flalign}
 &2BB'\tau  + 2B^2 \eta _y  - B^2 \mu _s
- 2B^2 x\varphi _y  = 0,&\label{e62}
\end{flalign}
\begin{flalign}
 &2B^2 x~\xi  + 2BB'x^2~ \tau  + 2CC'\tau  - 2B^2 x~\eta~ _z
 \, - C^2 \mu _s  - B^2 x^2 \mu _s  + 2C^2 \varphi _z  + 2B^2 x^2 \varphi _z  =
0,&\label{e63}
\end{flalign}
\begin{flalign}
&A^2~ \xi~ _t  - \tau~ _x  = 0,&\label{e64}\\
&B^2 ~\eta~ _t  - \tau ~_y  - B^2 x~\varphi~ _t  = 0,&
\label{e65}\\
 &- B^2 x~\eta _t  - \tau~ _z  + C^2~
\varphi~ _t  + B^2 x^2~ \varphi~ _t = 0,&\label{e66}\\
 &B^2~ \eta~ _x + A^2~ \xi~ _y  - B^2 x~\varphi~ _x  = 0,&
\label{e67}\\
& - B^2 x~\eta~ _x  + A^2~ \xi~ _z  + C^2 ~\varphi~ _x + B^2~ x^2
\varphi~ _x  = 0,&\label{e68}
\end{flalign}
\begin{flalign}
&- B^2 ~\xi  - 2BB'~x~\tau  - B^2 x~\eta~ _y  + B^2 ~\eta~ _z  +
 B^2~ x\mu~ _s  + C^2~ \varphi~ _y  + B^2 ~x^2~ \varphi~ _y  - B^2~ x~\varphi~ _z  =
 0,&\label{e69}
\end{flalign}
\begin{flalign}
&f~_t + 2~\tau~ _s  = 0,&\label{e70}\\
& - f~_x  + 2A^2~ \xi~ _s  = 0,&\label{e71}\\
& - {\rm{ }}f~_y  + 2B^2~ \eta~ _s  - 2B^2 ~x\varphi~ _s  =
0,&\label{e72}\\
& - f~_z  - 2B^2~ x~\eta _s  + 2C^2~ \varphi~ _s  + 2B^2~ x^2
~\varphi _s  = 0,&\label{e73}\\
&f~_s = 0.&\label{e74}
\end{flalign}
In order to classify Noether symmetries, the aim is to determine
coefficient of the Noether's  operator $'X'$ as functions of $t,
x, y, z$ and $s$. The method adopted to classify Noether
symmetries admitted by (\ref {metric})~is to solve the  system
(\ref {e56})~-~(\ref {e74})~by following an exhaustive procedure
followed in \cite {ccs}. This method of solving above system gives
rise to Noether symmetries admitted by (\ref {metric})
corresponding to allowable conditions satisfied by conditions of
the metric components. This process of classification gives a
total of ten cases. Each case satisfies certain differential
constraints given in equation (\ref {metric}) as detailed below .

{\bf Case I.}~~$A''=0,~A=B=C$.\\
The component form of the Noether symmetry $X=\mu \frac{\partial
}{{\partial s}} + \tau \frac{\partial }{{\partial t}} + \xi
\frac{\partial }{{\partial x}} + \eta \frac{\partial }{{\partial
y}} + \varphi \frac{\partial
}{{\partial z}} ~$are:\\
$\mu  = {\alpha _1\over 2} s^2 + \alpha _2 s + \alpha _3
 ;\tau= {\alpha _1\over 2}s~t + {\alpha _2\over 2}t + \alpha
_4 s + \alpha _6;\xi= \alpha_{7}; \eta  = \alpha _7 z + \alpha
~_8; \phi  =  -\alpha_{9}$, along with conformal factor $f= -
{\alpha _1\over 2}~t^2 - 2\alpha _4~ t.$\\
The Noether symmetry generators in this case are:\\
$X_1  = \frac{{s^2 }}{2}\frac{\partial }{{\partial s}} +
\frac{{ts}}{2}\frac{\partial }{{\partial t}}, f =  - \frac{{t^2
}}{2};~X_2  = s\frac{\partial }{{\partial s}} +
\frac{t}{2}\frac{\partial }{{\partial t}};~X_3  = \frac{\partial
}{{\partial t}};~X_4  = \frac{\partial }{{\partial s}},X_5  =
s\frac{\partial }{{\partial t}},~f =  - 2t;~ X_6  = \frac{\partial
}{{\partial x}} + z\frac{\partial }{{\partial y}};\\X_7  =
\frac{\partial }{{\partial y}};~X_8  =  - \frac{\partial
}{{\partial z}}.$ \\
The non-commuting generators satisfy the algebra given by:\\
$[X_1,X_2]=-X_1,~[X_1,X_3]={1\over
2}~X_5,~[X_1,X_4]=-~X_2,~[X_2,X_3]=-{1\over
2}~X_1,~[X_2,X_4]=-~X_4, [X_2,X_5]=~X_5,$\\ $ [X_4,X_5]=X_3,
[X_6,X_8]=X_7.$\\ All the remaining commutation relations are
zero.\\
{\bf Case II.}~~ $A'=0,~B'=0,~C'=0$; \\
 In this case the components of the Noether symmetry generator $X=\mu \frac{\partial }{{\partial s}} + \tau
\frac{\partial }{{\partial t}} + \xi \frac{\partial }{{\partial
x}} + \eta \frac{\partial }{{\partial y}} + \varphi \frac{\partial
}{{\partial z}} $~,~along with gauge term, are given by\\
$\mu=\alpha_1;\tau=\alpha_2~s+\alpha_3;~\xi=\alpha_4~z+\alpha_5;~\eta=\alpha_4({z^2-x^2\over2})+\alpha_5z+\alpha_6;\phi=-\alpha_4x-\alpha_7$
with $f=-2\alpha_2t$\\
In generator form, the above give rise to following 7 symmetry generators:\\
 ~~~$X_1  = \frac{\partial }{{\partial s}},~X_2  = s\frac{\partial }{{\partial t}}~with
f=-2t;~
 X_3  = z\frac{\partial }{{\partial x}} +({z^2-x^2\over 2}){\partial \over \partial y} - x\frac{\partial }{{\partial
 z}},~ X_4  = \frac{\partial }{{\partial t}},~X_5  = \frac{\partial }{{\partial x}} + z\frac{\partial }{{\partial y}},~X_6  = \frac{\partial }{{\partial y}},~X_7  =  - \frac{\partial }{{\partial
z}}$\\
The above symmetries form a closed form Lie-algebra of which all
generators commute except\\
$\[X_1,X_2\]=X_4; \[X_5,X_3\]=X_7;\[X_3,X_7\]=X_5 ;\[X_5,X_7\]=X_6.$\\
{\bf Case III.}~~$A'=0,~B'=0,~C''=0$; \\
The components of the Noether symmetry generator $X=\mu
\frac{\partial }{{\partial s}} + \tau \frac{\partial }{{\partial
t}} + \xi \frac{\partial }{{\partial x}} + \eta \frac{\partial
}{{\partial y}} + \varphi \frac{\partial }{{\partial z}} $~ are:\\
$\mu=\alpha_1~s+\alpha_2;\tau={\alpha_1\over
2}{t}+\alpha_3;\xi={\alpha_1\over
2}x+\alpha_4;\eta={\alpha_1\over 2}y+\alpha_4z+\alpha_5;\phi=-\alpha_6$\\
{Generator form of above Noether symmetries is:}\\
 $X_1  = s\frac{\partial }{{\partial s}} + \frac{t}{2}\frac{\partial }{{\partial t}} + \frac{x}{2}\frac{\partial }{{\partial x}} + \frac{y}{2}\frac{\partial }{{\partial y}},~X_2  = \frac{\partial }{{\partial s}},~ X_3  = \frac{\partial }{{\partial t}},~X_4  = \frac{\partial }{{\partial x}} + z\frac{\partial }{{\partial y}},~X_5  = \frac{\partial }{{\partial y}},~X_6  =  - \frac{\partial }{{\partial
z}}$\\
{\bf Case IV.}~~$A'=0,~B'\neq0,~C'=0$.\\
In this case the components of Noether symmetry generator $X=\mu
\frac{\partial }{{\partial s}} + \tau \frac{\partial }{{\partial
t}} + \xi \frac{\partial }{{\partial x}} + \eta \frac{\partial
}{{\partial y}} + \varphi \frac{\partial
}{{\partial z}} $~ take the form,\\
$\mu=\alpha_1; \tau=0;
\xi=\alpha_2z+\alpha_3;\eta=\alpha_2({z^2-x^2\over
2})+\alpha_{3}z+\alpha_4;\phi=-\alpha_{3}x-\alpha_5.$\\
Noether Symmetries generators associated with above components are:\\
$X_1  = \frac{\partial }{{\partial s}},~X_2  = z\frac{\partial
}{{\partial x}} - \frac{{x^2 }}{2}\frac{\partial }{{\partial y}} +
\frac{{z^2 }}{2}\frac{\partial }{{\partial y}} - x\frac{\partial
}{{\partial z}},~
 X_3  = \frac{\partial }{{\partial x}} + z\frac{\partial }{{\partial y}},~X_4  = \frac{\partial }{{\partial y}},~X_5  =  - \frac{\partial }{{\partial z}}. \\$
Except~ $[X_2,X_5]=X_3, [X_3,X_2]=X_5, [X_3, X_5]=X_4$ , all the
remaining commutation relations vanish. \\
{\bf Case V.}~~$A''\neq0,~B''=0=C''$.\\
Non-zero components of the Noether symmetry generator, $X=\mu
\frac{\partial }{{\partial s}} + \tau \frac{\partial }{{\partial
t}} + \xi \frac{\partial }{{\partial x}} + \eta \frac{\partial
}{{\partial y}} + \varphi \frac{\partial
}{{\partial z}} $~ become,\\
$\mu=\alpha_1s+\alpha_2; \tau={\alpha_1 \over
2}t+\alpha_3;\eta=\alpha_4;\phi=-\alpha_5.$\\
Generator form of above symmetries is given by:\\
$X_1  = s\frac{\partial }{{\partial s}} +
\frac{t}{2}\frac{\partial }{{\partial t}};~X_2  = \frac{\partial
}{{\partial s}};~ X_3  = \frac{\partial }{{\partial t}};~X_4  =
\frac{\partial }{{\partial y}};~X_5  =  - \frac{\partial
}{{\partial z}}.$\\
Two non-commuting relations are $[X_2, X_1]=X_2,
[X_3,X_1]={-1\over 2}~X_3.$\\
{\bf Case VI.}~~$A''=0,~B'=0,C=C(t)$.\\
In this case $C$ is an arbitrary function of $t$ and the non-zero components of the Noether symmetry generator take the form,\\ $\mu, \tau, \eta, \phi$ with $\xi=0$.\\
The five Noether symmetry generators associated with above
generator are;\\
$\mu  = \alpha _1 s + \alpha _2; \tau= \alpha _1 (\frac{t}{2} +
1); \eta  = \alpha _3 y + \alpha _4 ;\phi  =  \alpha _3 z -\alpha_{5}$.\\
In the light of above, the symmetry generators take the form,\\
$X_1  = s\frac{\partial }{{\partial s}} + (\frac{t}{2} +
1)\frac{\partial }{{\partial t}},~X_2  = \frac{\partial
}{{\partial s}}, ~ X_3  = y\frac{\partial }{{\partial y}} +
z\frac{\partial }{{\partial z}},~X_4  = \frac{\partial }{{\partial
y}},~X_5  =  - \frac{\partial }{{\partial z}}. \\$ The
non-commuting commutation relation satisfied by these symmetries
are: $[X_2,X_1]=X_2,
[X_4,X_3]=X_4, [X_5,X_3]=X_5.$\\
{\bf Case VII.}~~$A'=0,~B'=0,~C''\neq0$;\\
Here the components of Noether symmetry generator $X=\mu
\frac{\partial }{{\partial s}} + \tau \frac{\partial }{{\partial
t}} + \xi \frac{\partial }{{\partial x}} + \eta \frac{\partial
}{{\partial y}} + \varphi \frac{\partial }{{\partial z}} $~ are,\\
 $\mu=\alpha_1; \tau=0; \xi=\alpha_2; \eta=\alpha_2z+\alpha_3;
\phi=-\alpha_4.$
In generator form, the above give rise to following 4 symmetry generators:\\
$ X_1  = \frac{\partial }{{\partial s}},X_2  = \frac{\partial
}{{\partial x}} + z\frac{\partial }{{\partial y}},X_3  =
\frac{\partial }{{\partial y}},X_4  =  - \frac{\partial
}{{\partial z}}.$ \\
All of the symmetry generators commute except $\[X_2,X_4\]=x_3.$\\
{\bf Case VIII.}~~$A''\neq0,~B'=0=C''$.\\
The non zero components of the Noether symmetry generator, $X=\mu
\frac{\partial }{{\partial s}} + \tau \frac{\partial }{{\partial
t}} + \xi \frac{\partial }{{\partial x}} + \eta \frac{\partial
}{{\partial y}} + \varphi \frac{\partial
}{{\partial z}} ~$~are,\\
$\mu=\alpha_1;\tau=\alpha_2;\eta=\alpha_3;\phi=-\alpha_4.$\\Noether
symmetry generator associated with above components are:\\
$X_1  = \frac{\partial }{{\partial s}},~X_2  = \frac{\partial
}{{\partial t}};~X_3  = \frac{\partial }{{\partial y}};~X_4  =  -
\frac{\partial }{{\partial z}}.$\\
All Noether symmetry generators in this case commute, i.e., $[X_i,X_j]=0$ for all $i$ and $j$.\\
{\bf Case IX.}~~$A''\neq0,~B=B(t),~C''=0$.\\
In this case we get minimal set of 3 Noether symmetries, which in component form are given by:\\
$\mu=\alpha_1; \eta=\alpha_2;, \phi=-\alpha_3$.\\
The three Noether symmetry generators corresponding to above
components are;\\
$X_1=\frac{\partial }{{\partial s}};~ X_2=\frac{\partial
}{{\partial y}};~X_3=\frac{\partial }{{\partial z}},~$all of which commute.\\
We do not include the remaining cases in this discussion because
they either belong to one of the above cases or are their
proper subalgebras.\\
{\bf Structure of Lie Algebras of Noether Symmetries}\\ We give
the detailed structure of the Lie algebra of Noether symmetries in
two cases, ({\bf {I}}) and ({\bf {II}}). The main reason for this
is that the Lie algebra in {\bf case I} is non-soluble, while it
is soluble in all the remaining cases, which can be dealt with as case ({\bf case II}). \\
{\bf Structure of the Lie Algebra in Case I}\\
The Lie algebra of Noether symmetries in this case is $~L_8  = <
X_1 ,X_2 ,X_3 ,X_4,X_5,X_6,X_7,X_8 >.~$ The Killing form is
defined by  $~\kappa (X_i ,X_j )=Tr(Ad X_i~o~ Ad X_j).$  It is
straight forward to verify that $\kappa(X_1 ,X_4 )$ and
$\kappa(X_2 ,X_2 ) $ are not equal
to zero, while all the remaining $~\kappa(X_i ,X_j ) = 0.$ The Levi decomposition is $<X_1,X_2,X_4>\oplus Rad(L_1)$, where the first summand is isomorphic to $sl(2,R).$\\
{\bf Structure of the Lie Algebra in Case II}\\
The Lie algebra of Noether symmetries in this case is , $L_1  = <
X_1 ,X_2 ,X_3 ,X_4 ,X_5 ,X_6 ,X_7 >$. A straightforward
computation shows that the derived series of $L$ is $L\supseteq
L^{(1)}\supseteq L^{(2)}\supseteq L^{3}=\{0\}.$  This can be used to give an ascending sequence of ideals each of codimension 1, which is useful in obtaining conjugacy
 classes of 1$-$dimensional subalgebra as in {\cite{ibragimov}.\\
{\bf Summary and Discussions}\\
A classification of Noether symmetries of a Lagrangian associated
with the Bianchi type II spacetime metric is obtained. This
classification arises from a variety of differential constraints
obtained on the coefficients of the metric representing the
Bianchi models. It is shown that for a total of 9 nontrivial cases
arise giving Noether symmetries admitting maximal group $G<8>$ and
a minimal group $G<3>$ of Noether symmetries. The maximal group
$G<8>$ contains all the remaining seven to three parameter
subalgebras as proper subalgebra.\\ All these cases yield
physically interesting conservation laws giving linear momentum
conservation along $y$ and $z$ directions, whereas cases $I-III,
IV$ and $VIII$ additionally admit time translational invariance
giving conservation of energy. \\ The detailed structure of the
algebras given above can be used to construct conserved quantities
for each conjugacy classes of 1$-$dimensional subalgebras as in
(\cite {ibragimov, chamatchot}).

\end{document}